\documentclass[runningheads]{llncs}
\usepackage[T1]{fontenc}
\usepackage{caption}
\usepackage{subcaption}
\usepackage{amsmath,amsfonts}
\usepackage{algorithm}
\usepackage{algpseudocode}
\usepackage{array}
\usepackage[caption=false,font=normalsize,labelfont=sf,textfont=sf]{subfig}
\usepackage{textcomp}
\usepackage{stfloats}
\usepackage{url}
\usepackage{verbatim}
\usepackage{graphicx}
\usepackage{amssymb}
\usepackage{multirow}
\usepackage[table]{xcolor}
\definecolor{mycolor}{HTML}{FFFC9E}
\newcolumntype{a}{>{\columncolor{mycolor}}c}
\usepackage{cite}
\definecolor{mycolor2}{HTML}{9AFF99}
\newcolumntype{b}{>{\columncolor{mycolor2}}c}
\usepackage{xcolor}

\newcommand{\LOOPPE}{$\mathcal{LOOP-PE}$}

\newcommand{\LOOPLCP}{$\mathcal{LOOP-LC}\hspace{.1cm}2.0$}
\usepackage{graphicx}
\begin{document}
\title{Towards Reliable Neural Optimizers: A Permutation Equivariant Neural Approximation for Information Processing Applications}
%
%
\author{Meiyi Li\inst{1}\orcidID{0000-0002-0178-7883} \and
Javad Mohammadi\inst{1}\orcidID{0000-0003-0425-5302} }
\authorrunning{Meiyi Li,  Javad Mohammadi.}
%
\institute{The University of Texas at Austin,  TX , USA \\
\email{\{meiyil,javadm\}@utexas.edu}}
\maketitle              
\vspace{-0.3in}
\begin{abstract}
The complexities of information processing across Dynamic Data Driven Applications Systems drive the development and adoption of Artificial Intelligence-based optimization solutions. Traditional solvers often suffer from slow response times and an inability to adapt swiftly to real-time input variations. To address these deficiencies, we will expand on our previous research in neural-based optimizers by introducing a machine learning-enabled neural approximation model called \LOOPPE~ (Learning to Optimize the Optimization Process -- Permutation Equivariance version). This model not only enhances decision-making efficiency but also dynamically adapts to variations of data collections from sensor networks. In this work, we focus on mitigating the heterogeneity issues of data collection from sensor networks, including sensor dropout and failures, communication delays, and the complexities involved in integrating new sensors during system scaling. The proposed \LOOPPE~ model specifically overcomes these issues with a unique structure that is permutation equivariant, allowing it to accommodate inputs from a varying number of sensors and directly linking these inputs to their optimal operational outputs. This design significantly boosts the system's flexibility and adaptability, especially in scenarios characterized by unordered, distributed, and asynchronous data collections. Moreover, our approach increases the robustness of decision-making by integrating physical constraints through the generalized gauge map method, which theoretically ensures the decisions' practical feasibility and operational viability under dynamic conditions. We use a DDDAS case study to demonstrate that \LOOPPE~ model reliably delivers near-optimal and adaptable solutions, significantly outperforming traditional methods in managing the complexities of multi-sensor environments for real-time deployments.

\end{abstract}
\vspace{-0.4in}
\section{Introduction}
\vspace{-0.1in}

Sensors are crucial for Dynamic Data Driven Applications Systems (DDDASs) \cite{ref_article_0} as providing continuous and accurate data feed is essential for instantaneous decision-making and system optimization. Sensor data integration boosts operational efficiency and adaptability, facilitating effective management in complex, time-sensitive environments.

At the heart of dynamic management within these sensor-based systems lies the frequent need to solve large-scale economic optimization problems quickly \cite{ref_article_1}. Traditional iterative solvers are limited by extended computation times in time-sensitive scenarios. In contrast, recent advancements in machine learning (ML) streamline the optimization process significantly \cite{ref_article_2}. These neural approximators leverage abundant historical data in offline computations to reduce the iterations needed to reach optimal solutions, enhancing speed and efficiency \cite{ref_article_3,ref_article_4}.

The information processing in sensor-based systems, however, is complicated by factors such as sensor dropout due to technological failures, transmission delays, and complications arising from the integration of new sensors during system expansion. These challenges necessitate adaptive ML models that can dynamically accommodate changes and maintain reliable, viable solutions under time-varying system constraints \cite{ref_article_5}. Conventional models, with their fixed input dimensions, often fail to handle unordered, distributed, and asynchronous data collections. These are common challenges in environments that are both contested and congested. This situation underscores the need for more flexible approaches that allow for seamless integration or removal of sensors without disrupting ongoing operations \cite{ref_article_5.1}, \cite{ref_article_5.2}.

A critical challenge in applying ML to sensor system management is ensuring the practical feasibility of solutions, particularly compliance with necessary physical and engineering constraints. Traditional methods involving simple penalty terms \cite{ref_article_6} or projection methods \cite{ref_article_7} for refining solutions impose soft constraints or require additional iterative processing, which is limiting for real-time applications. Our prior work \cite{ref_article_8} utilizes gauge map functions in a non-iterative, feed-forward approach to strictly adhere to linear constraints, thus minimizing computational demands.

This paper presents an ML-enabled neural approximator named $\mathcal{LOOP-PE}$ (Learning to Optimize the Optimization Process -- Permutation Equivariance version), designed to optimize operations in dynamically changing sensor networks. Our model's unique permutation equivariant structure allows it to process inputs from an indefinite number of sensors and directly map these inputs to their optimal operational outputs, which largely enhances the flexibility and adaptability of information processing in DDDAS. Further, our model extends our previous work \cite{ref_article_8} which integrates physical constraints through the generalized gauge map method to improve the robustness of decision-making. This integration theoretically ensures the decisions' practical feasibility and operational viability under dynamic conditions. Comprehensive case studies demonstrate that our \LOOPPE~ model delivers reliable, near-optimal, and adaptable solutions, significantly outperforming traditional methods in managing the complexities of multi-sensor environments for real-time decision-making.

\vspace{-0.2in}

\section{Problem formulation} 
\vspace{-0.15in}
In this paper, we consider a dynamic sensor network, represented by a set of sensors denoted as $\mathcal{N}_{\texttt{A}}$, where each sensor is indexed by $i$, such that $i \in \mathcal{N}_{\texttt{A}}$. The roles and functionalities of sensors in this network can adapt in response to internal reconfigurations and external environmental changes. To manage this dynamic nature effectively, our system is designed to be inherently indifferent to the permutation of sensor indices, ensuring that each sensor is treated equivalently under any permutation.

\vspace{-0.1in}
\subsection{ Permutation Equivariance}
\vspace{-0.1in}

The foundation of our approach is permutation equivariance, which is essential for maintaining the system's effectiveness and consistency. Permutation equivariance ensures that the outcomes of our optimization process remain invariant regardless of the order of sensor indices. This attribute allows the system to dynamically adapt to changing sensor roles without losing functionality or efficiency. In mathematical terms, for any permutation $\sigma$ over the set $\mathcal{N}_{\texttt{A}}$, the optimizer satisfies:
\small
\vspace{-0.1in}
\begin{align}
    \left [ \mathbf{u}^{\sigma(1)},...,\mathbf{u}^{\sigma(i)},...,\forall i\in \mathcal{N}_{\texttt{A}}\right ]=\xi(\left [ \mathbf{x}^{\sigma(1)},...,\mathbf{x}^{\sigma(i)},...,\forall i\in \mathcal{N}_{\texttt{A}} \right ])\label{permutation}
\end{align}
\normalsize

\vspace{-0.3in}

\subsection{Objective Function and Constraints}

The primary objective in managing this sensor-based system is to optimize the collective behavior of all sensors while adhering to operational constraints. These constraints are twofold: (i) local constraints, which apply individually to each sensor dictating feasible solutions based on the sensor's own capabilities and information, and (ii) coupled constraints, which necessitate coordination among sensors as they depend on the collective actions or states of multiple sensors. Formally, the optimization problem is structured as follows:
\small
\vspace{-0.1in}
\begin{subequations}
    \label{distributed compact}
\begin{gather}
    \min  
    f(\mathbf{u},\mathbf{x})=\sum_{i\in \mathcal{N}_{\texttt{A}}}f^i(\mathbf{u}^i,\mathbf{x}^i) \label{distributed compact objective}\\
     \mathbf{u}=\left [ \mathbf{u}^1,...,\mathbf{u}^i,..., \forall i\in \mathcal{N}_{\texttt{A}}  \right ], \mathbf{x}=\left [ \mathbf{x}^1,...,\mathbf{x}^i,..., \forall i\in \mathcal{N}_{\texttt{A}} \right ]  \\
\textup{local constraints: }\left\{\begin{matrix}
\mathbf{A}_{\texttt{eq}}(\mathbf{x}^i)\mathbf{u}^i+\mathbf{B}_{\texttt{eq}}(\mathbf{x}^i)=\mathbf{0}\\ 
\mathbf{A}_{\texttt{ineq}}(\mathbf{x}^i)\mathbf{u}^i+\mathbf{B}_{\texttt{ineq}}(\mathbf{x}^i)\leq\mathbf{0}
\end{matrix}\right. , \forall i\in \mathcal{N}_{\texttt{A}} \label{distributed compact local}\\
\textup{coupled constraints: } \sum_{i\in \mathcal{N}_{\texttt{A}}}[\mathbf{A}(\mathbf{x}^i)\mathbf{u}^i+\mathbf{B}(\mathbf{x}^i)]\leq \mathbf{0}\label{distributed compact connection}
\end{gather}
\vspace{-0.2in}
\end{subequations}
\normalsize

Here, $\mathbf{u}$ represents the collection of optimization variables for all sensors, and $\mathbf{u}^i$ is the optimization variable vector specific to sensor $i$. Similarly, $\mathbf{x}$ includes all input parameters across sensors, with $\mathbf{x}^i$ denoting the input parameter vector for sensor $i$. The function $f$ encapsulates the overall objective by summing up objective functions $f^i$. Equations \eqref{distributed compact local} and \eqref{distributed compact connection} specify the local and coupled constraints, respectively, with $\mathbf{A}_{\texttt{eq}}$, $\mathbf{B}_{\texttt{eq}}$, $\mathbf{A}_{\texttt{ineq}}$, and $\mathbf{B}_{\texttt{ineq}}$ detailing the forms these constraints take for each sensor. $\mathbf{A}$,
and $\mathbf{B}$  capture the compact form of coupled constraints among all sensors. They are expressed with $(\mathbf{x}^i)$ because they are determined by $\mathbf{x}^i$. Also, $\mathbf{0}$ symbolizes an all-zero vector.

We use $\mathbf{u}^*$ to represent the optimal solution to problem \eqref{distributed compact}. Our real-time management model aims to provide an optimizer $\xi$ that maps these input parameters $\mathbf{x}$ to the optimal solution $\mathbf{u}^*$. In the subsequent sections, we will introduce a neural network architecture designed to exploit this permutation-equivariant framework to address the optimization challenges described.

\vspace{-0.1in}

\begin{figure}[h]
\centering
\includegraphics[width=0.8\textwidth]{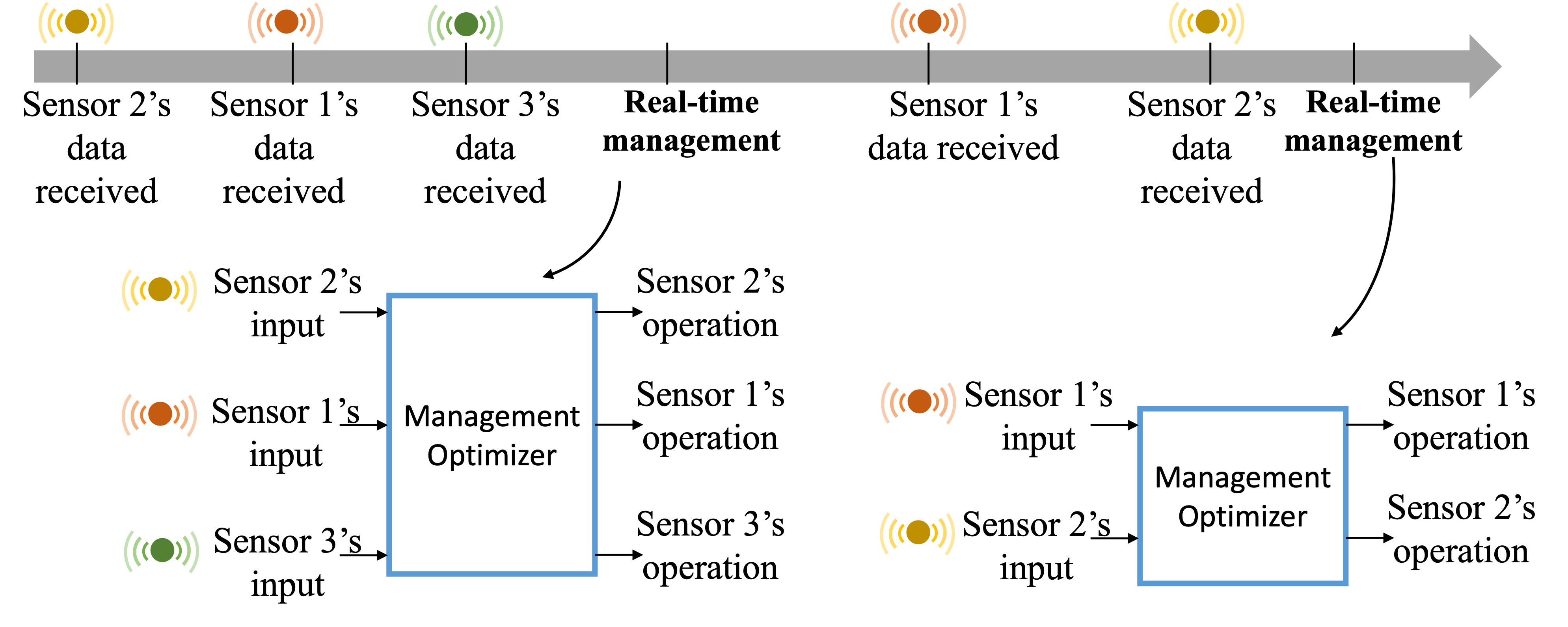}
\vspace{-0.15in}
\caption{Illustration of the Permutation Equivariance Property in the Real-Time Management Optimizer. This feature ensures output corresponds to input order, enabling adaptability to sensor addition or removal without extensive reconfiguration. }
\label{example}
\vspace{-0.2in}
\end{figure}

\vspace{-0.1in}
\section{Proposed Method}
\begin{figure}[h]
\centering
\includegraphics[width=0.9\textwidth]{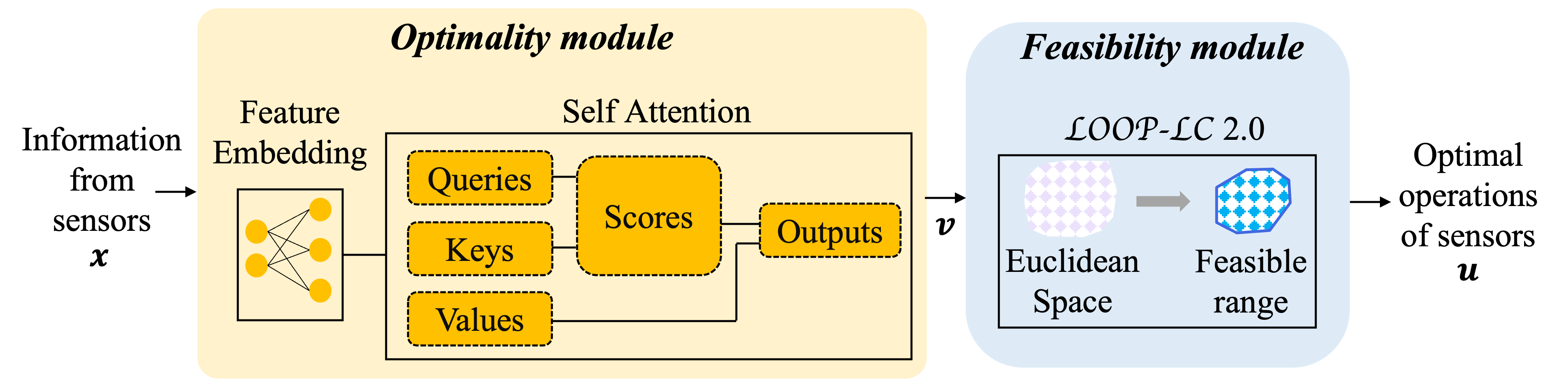}
\vspace{-0.1in}
\caption{Building blocks of the proposed \LOOPPE~ model. The Optimality Module uses an attention mechanism to process input from varying sensor numbers and generate virtual predictions. The Feasibility Module uses the \LOOPLCP~ model \cite{ref_article_8} to convert these predictions into practical, constraint-compliant actions, ensuring flexibility and robustness across different sensor setups and dynamics.}
\label{block}
\vspace{-0.3in}
\end{figure}

\vspace{-0.1in}
In this section, we introduce a permutation-equivariant neural approximator \LOOPPE~,  designed to tackle the optimization problem detailed in \eqref{distributed compact}, specifically within a dynamically changing sensor network. Figure \ref{block} illustrates the core components of our method, which include the Optimality and Feasibility Modules. 

\vspace{-0.3in}
\vspace{-0.2in}
\subsection{Optimality Module}

The optimality module consumes the data from sensors ($\mathbf{x}$) and output virtual predictions for each($\mathbf{v}$), it consists of two submodules: 

(i) Feature Embedding: Use a fully connected neural network to project the input features of each sensor to a higher dimensional feature space. This is to capture more complex interactions in the subsequent attention mechanism.

(ii) Self Attention: Allow the network to consider and appropriately weigh the importance of each sensor's features relative to others. This self-attention mechanism is particularly beneficial for data where the interaction between sensors plays a more crucial role than their absolute positions in the input sequence.

\vspace{-0.15in}
\subsection{Feasibility module}

The feasibility module utilizes our previous \LOOPLCP~ model \cite{ref_article_8} to convert the predictions $\mathbf{v}$ into practical actions $\mathbf{u}$ that adhere to both local and coupled constraints, maintaining the permutation-equivariance property throughout. 

The \LOOPLCP~ model is capable of mapping the Optimality module's solution to a feasible point within the linearly-constrained
domain.  For equality constraints in \eqref{distributed compact local}, it applies variable elimination to reduce the size of the problem to a reformulated problem only with independent variables. For inequality constraints in \eqref{distributed compact local} and \eqref{distributed compact connection}, a generalized gauge map is adopted to rescale any infeasible solutions to the boundary of the constraint set. The generalized gauge map will keep
the virtual predictions as they are if they have already lied
within the desired feasible range. Otherwise, it will rescale the virtual predictions to the
boundary. The feasibility module, whose mapping is denoted by $\mathbb{T}$ has the form below:
\small
\vspace{-0.15in}
\begin{align}
    \mathbf{u}=\mathbb{T}(\mathbf{v})=\mathbf{u}_{\texttt{0}}(\mathbf{x})+\frac{1}{\underset{r}{\max}\left \{ {\left [\frac{\sum_{i\in \mathcal{N}_{\texttt{A}}}{\mathbf{H}(\mathbf{x}^i)}\mathbf{v}^i}{\sum_{i\in \mathcal{N}_{\texttt{A}}}{\mathbf{h}(\mathbf{x}^i)}}  \right ]}^r  \right \}}\mathbf{v}\label{T}
\end{align}
\normalsize
\vspace{-0.15in}

where the matrix \(\mathbf{H}\) depends on the input data \(\mathbf{x}^i\) for each sensor, and the sum \(\sum_{i \in \mathcal{N}_{\texttt{A}}} \mathbf{H}(\mathbf{x}^i) \mathbf{v}^i\) yields a column vector. Similarly, the vector \(\mathbf{h}\) depends on \(\mathbf{x}^i\), and the sum \(\sum_{i \in \mathcal{N}_{\texttt{A}}} \mathbf{h}(\mathbf{x}^i)\) is also a column vector. The term \(\sum \mathbf{H}^i \mathbf{v}^i / \sum \mathbf{h}^i\) represents element-wise division, producing a column vector. The notation \([...]^r\) identifies the maximum value across all row elements. \(\mathbf{u}_{\texttt{0}}\) is an interior point within the feasible range, and it is permutation equivariant on \(\mathbf{x}\). Therefore, \(\mathbb{T}\) is a predefined function with a closed-form representation. The detailed expression is available in \cite{ref_article_8}.

\vspace{-0.2in}
\subsection{Invariance to Input Order}
\vspace{-0.1in}

The Optimality Module, denoted by \(\mathbb{O}\), achieves permutation equivariance because its feature embedding mechanism processes all features simultaneously, weighing them according to content rather than order. Its self-attention mechanism assesses the relationship between each sensor's features \(\mathbf{x}^i\) relative to others, independent of their sequence in the input.  Mathematically, for any permutation \(\sigma\), the output virtual predictions are:
\small
\[
\left[ \mathbf{v}^{\sigma(1)},  \ldots, \mathbf{v}^{\sigma(i)}, \ldots,\forall i\in \mathcal{N}_{\texttt{A}}  \right] = \mathbb{O}\left( \left[ \mathbf{x}^{\sigma(1)},  \ldots, \mathbf{x}^{\sigma(i)}, \ldots,\forall i\in \mathcal{N}_{\texttt{A}}  \right] \right)
\]
\normalsize

Here, \(\mathbb{O}\) maps input data \(\mathbf{x}\) to predictions \(\mathbf{v}\), with permutation \(\sigma\) ensuring that input and output orders are consistent.

The Feasibility Module, denoted by \(\mathbb{T}\), also maintains permutation equivariance of input \(\mathbf{v}\) and output \(\mathbf{u}\). The function \(\mathbf{u}_{\texttt{0}}\) is permutation equivariant on the input data \(\mathbf{x}\), and \(\mathbf{v}\) follows the same rule. The denominator term, computing the maximum across permutations, remains invariant to any permutation of \(\mathbf{v}\). Thus, \(\mathbb{T}\) ensures:

\small
\[
\left [ \mathbf{u}^{\sigma(1)}, \ldots,\mathbf{u}^{\sigma(i)}, \ldots, \forall i\in \mathcal{N}_{\texttt{A}}\right ] = \mathbb{T}\left( \left[ \mathbf{v}^{\sigma(1)}, \ldots,\mathbf{v}^{\sigma(i)}, \ldots, \forall i\in \mathcal{N}_{\texttt{A}} \right] \right)
\]
\normalsize

Combined, the permutation-equivariant mappings \(\mathbb{O}\) and \(\mathbb{T}\) maintain permutation equivariance, enabling the \LOOPPE~ architecture to handle unordered, varying numbers of inputs while preserving consistent relationships across changing sensor setups.

\vspace{-0.2in}
\section{ Experiment results}
\vspace{-0.1in}
We implemented our proposed method in a case study involving a Virtual Power Plant (VPP) tasked with managing the assets of 20 distributed generator agents, collectively referred to as $\mathcal{N}_{\texttt{A}}$. These agents were integrated and coordinated with power grid operations to optimize the use of distributed energy. The primary objective of real-time energy management within the VPP is to maximize the utilization of distributed energy while adhering to generation limits and power system constraints. The optimization model is formulated as follows:

\vspace{-0.1in}
\small
\begin{subequations}
    \label{vpp}
    \begin{gather}
        \min \sum_{i \in \mathcal{N}_{\texttt{A}}} \left( P_{\texttt{G}}^i - P_{\texttt{C}}^i \right)^2
        \label{vpp_objective} \\
        0 \leq P_{\texttt{G}}^i \leq P_{\texttt{C}}^i, \forall i \in \mathcal{N}_{\texttt{A}}
        \label{vpp_generation} \\
        -P_{\texttt{omax}} \leq \sum_{i \in \mathcal{N}_{\texttt{A}}} (P_{\texttt{G}}^i - P_{\texttt{D}}^i) \leq P_{\texttt{omax}}
        \label{vpp_system}
    \end{gather}
\end{subequations}
\normalsize
\vspace{-0.1in}

where $P_{\texttt{G}}^i$ represents the generation power of agent $i$, $P_{\texttt{C}}^i$ denotes the generation capability, and $P_{\texttt{D}}^i$ indicates the load demand. Equation \eqref{vpp_objective} penalizes the wasted resources. Equation \eqref{vpp_generation} ensures generation remains within the agents' capacity, and \eqref{vpp_system} confines the net output of the VPP within the threshold $P_{\texttt{omax}}$. Therefore, the input parameters $\mathbf{x}^i$ of the real-time optimizer are $[P_{\texttt{C}}^i,P_{\texttt{D}}^i]$ and the optimization variables $\mathbf{u}^i$ are $P_{\texttt{G}}^i$, $\forall i\in \mathcal{N}_{\texttt{A}}$.

We use the parameters in \cite{ref_article1} where the agent's generation capability varies from $10kW$ to $25kW$. Also, $P_{\texttt{omax}}=100 kW$. We account for a 10 \% fluctuation of each agent. For each sample, assume only a random number of sensors received the corresponding agent's input parameters. 400 samples are collected where 100
of them being designated as test data points. The well-known commercial solver GUROBI \cite{gurobi} is used as the baseline for comparison.

Table \ref{tab:time_results} presents the computational performance of the proposed method. The variability in sensor data contributes to fluctuations in the scale of the optimization problem, which in turn affects the search time for the optimal solution. Regardless of the metric considered, e.g., minimum, maximum, or average execution time, the proposed method consistently achieves significantly faster solution times compared with traditional solvers.
\vspace{-0.35in}
\begin{table}[htbp]
\caption{Computational Time Comparison}\centering
\begin{tabular}{|c|c|c|}
\hline
Performance Metric & Gurobi Solver Time (ms) & Proposed Method Time (ms) \\ \hline
Average            & 6.48                     & 0.33                        \\ \hline
Minimum            & 5.02                      & 0.30                      \\ \hline
Maximum            & 24.34                     & 0.58                      \\ \hline
\end{tabular}
\label{tab:time_results}
\end{table}
\vspace{-0.2in}

Table \ref{tab:gaps} displays the solution optimality gap and feasibility gap for our method, using the commercial solver GUROBI \cite{gurobi} as the baseline. The optimality gap measures how closely the operational profiles align with the optimal solution across different samples, calculated as $\frac{\left\|\mathbf{u} - \mathbf{u}^*\right\|^2}{\left\|\mathbf{u}^*\right\|^2}$. The feasibility gap quantifies any violations within the operational profiles, reflecting their compliance with set constraints. 
Given our method's capability to provide feasible solutions, the solutions consistently exhibit a zero feasibility gap. This ensures the viability of the achieved solution under the dynamic operational conditions of DDDAS such as the power grid.

\begin{table}[htbp]
\vspace{-0.3in}
\caption{Optimality and Feasibility Gaps of Our \LOOPPE~ Method. }
\centering
\begin{tabular}{|c|ccc|c|}
\hline
\multirow{2}{*}{Metric} & \multicolumn{3}{c|}{Optimality Gap}                                   & Feasibility gap \\ \cline{2-5} 
                        & \multicolumn{1}{c|}{Average} & \multicolumn{1}{c|}{Minimum} & Maximum & Minimum         \\ \hline
Compared against baseline                   & \multicolumn{1}{c|}{0.04}    & \multicolumn{1}{c|}{0.00}    & 0.13    & 0.00            \\ \hline
\end{tabular}
\label{tab:gaps}
\vspace{-0.1in}
\end{table}

\begin{figure}[h]
\vspace{-0.2in}
\centering
\includegraphics[width=0.9\textwidth]{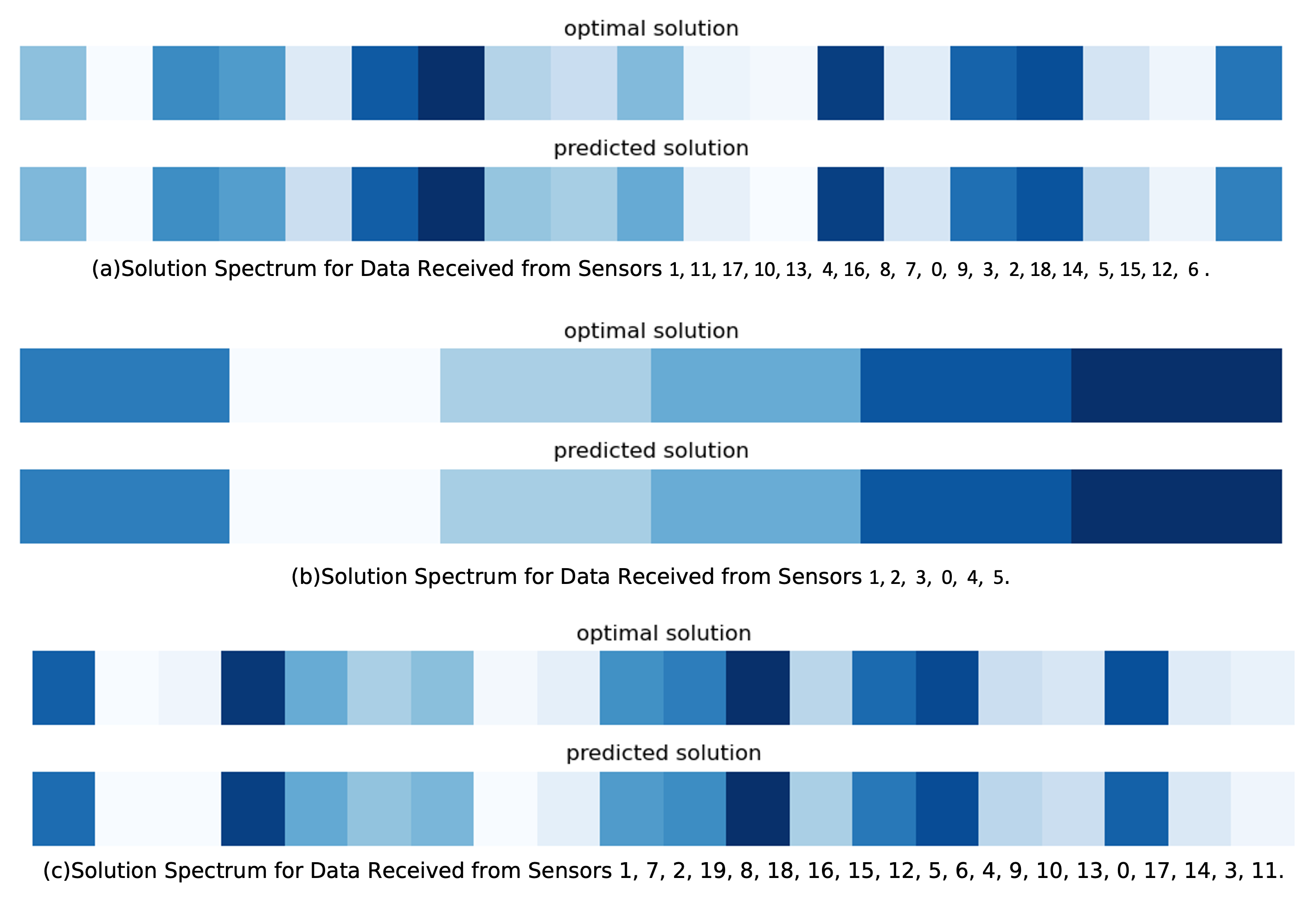}
\caption{Solution spectrum for various test samples using data from multiple sensors. This figure displays the solution spectra obtained from different sensors, illustrating the \LOOPPE~method's performance across diverse conditions. The ability to accept input in any order makes the model more adaptable to changes in the data collection setup, enhancing its flexibility and robustness in dynamic environments.}
\label{fig:spectrum_comparison}
\vspace{-0.4in}
\end{figure}

Figure \ref{fig:spectrum_comparison} compares the solution spectra of several test samples between the proposed method and traditional solvers. The results demonstrate that our method can find near-optimal solutions. In the figure, the similarity in the spectra between the solutions obtained using our method and those achieved with traditional solvers is clearly depicted.

\vspace{-0.15in}
\section{ Conclusion}
\vspace{-0.15in}
This paper introduces a permutation equivariant machine learning-enabled neural approximator \LOOPPE, designed to optimize operations while handling the complexities of information processing in DDDASs. The unique permutation equivariant structure of our model allows it to seamlessly process data from an indefinite number of sensors, mapping these inputs to their optimal operational outputs. This significantly enhances the flexibility and adaptability of dynamic sensor networks and increases the reliability of our neural approximator. Building upon our previous work \cite{ref_article_8}, which integrated physical constraints to ensure reliable decision-making, our model demonstrates its efficacy through a DDDAS case study. These studies showcase that our model consistently delivers reliable, near-optimal solutions and outperforms traditional methods in managing the complexities of multi-sensor environments for real-time operations.

In conclusion, we propose further research in three directions: enhancing robustness against data uncertainties, developing decentralized control mechanisms for improved scalability, and advancing techniques for integrating complex and heterogeneous data types. These directions aim to significantly refine and expand the capabilities of dynamic sensor networks.

\vspace{-0.2in}
\section{Acknowledgment}
\vspace{-0.1in}
This research is funded under AFOSR grants \#FA9550-24-1-0099 and FA9550-23-1-0203.

%
%
%
%

\end{document}